\documentclass[aps,twocolumn]{revtex4}

\usepackage{amsmath}
\usepackage{graphicx}

\usepackage{amssymb}
\usepackage{stmaryrd}

\usepackage{epsfig}

\usepackage{color}

\def\ov#1{\overline{#1}}

\def\vb#1{\mbox{\boldmath$#1$}}
\def\pd#1#2{\frac{\partial #1}{\partial #2}}

\def\bdot{\,\vb{\cdot}\,}
\def\btimes{\,\vb{\times}\,}

\def\cal#1{\mathcal{#1}}

\newcommand{\bc}{\begin{center}}
\newcommand{\ec}{\end{center}}
\newcommand{\bt}{\begin{tabbing}}
\newcommand{\et}{\end{tabbing}} 
\newcommand{\be}{\begin{eqnarray*}}
\newcommand{\ee}{\end{eqnarray*}}
\newcommand{\bs}{\begin{slide}}
\newcommand{\es}{\end{slide}}

\begin{document}

\begin{flushright}
November 14, 2014
\end{flushright}

\title{Compact formulas for bounce/transit averaging \\ in axisymmetric tokamak geometry}

\author{F.-X.~Duthoit$^{1}$, A.~J.~Brizard$^{2}$, and T.S.~Hahm$^{3}$}
\affiliation{$^{1}$SNU Division of Graduate Education for Sustainabilization of Foundation Energy, Seoul National University, Seoul 151-742, South Korea \\ $^{2}$Department of Physics, Saint Michael's College, Colchester, VT 05439, USA \\ $^{3}$Department of Nuclear Engineering, Seoul National University, Seoul 151-742, South Korea} 

\begin{abstract}
Compact formulas for bounce and transit orbit averaging of the fluctuation-amplitude eikonal factor in axisymmetric tokamak geometry, which is frequently encountered  in bounce-gyrokinetic description of microturbulence, are given in terms of the Jacobi elliptic functions and elliptic integrals. These formulas are readily applicable to the calculation of the neoclassical susceptibility in the framework of modern bounce-gyrokinetic theory. In the long-wavelength limit for axisymmetric electrostatic perturbations, we recover the expression for the Rosenbluth-Hinton residual zonal flow [Rosenbluth and Hinton, Phys.~Rev.~Lett.~{\bf 80}, 724 (1998)] accurately.

\end{abstract}

%\pacs{52.30.Gz, 52.65.Tt}

\maketitle

\section{Introduction}

Polarization effects have played a crucial role in the development of gyrokinetic theory \cite{Brizard_Hahm} and its applications in gyrokinetic particle simulations \cite{Lin,Garbet}. In the modern bounce-gyrokinetic theory, polarization results from the difference between the orbit-averaged position of a charged particle and its reduced position \cite{Brizard_2013}. Each dynamical-reduction step introduced in the derivation of reduced Vlasov-Maxwell equations \cite{Brizard_2008} yields a new contribution to the reduced polarization. Hence, bounce-gyrokinetic theory \cite{Fong_Hahm,Brizard_2000}, which results from the combination of the guiding-center/gyrocenter and the bounce-center/bounce-gyrocenter dynamical reductions, includes four different polarization contributions \cite{Wang_Hahm}. 

On the one hand, classical polarization effects, which are associated with the two-step guiding-center/gyrocenter dynamical reduction, arise from the difference between the gyro-averaged position of a charged particle and its guiding-center and gyrocenter positions. On the other hand, neoclassical polarization effects, which are associated with the two-step bounce-center/bounce-gyrocenter dynamical reduction, arise from the difference between the bounce/transit-averaged position of a charged particle and its bounce-center and bounce-gyrocenter positions. While gyroangle averaging is essentially a local process (due to the smallness of the gyroradius with respect to the background magnetic-field length-scale), the process of bounce/transit-angle averaging is a non-local one (especially for trapped-particle guiding-center orbits) \cite{Fong_Hahm}. 

The purpose of the present paper is to present a compact formulation for bounce/transit averaging of the fluctuation-amplitude eikonal factor that relies on the representation of the trapped/passing-particle guiding-center orbits in simple axisymmetric tokamak geometry \cite{Brizard_2011,Brizard_2014} based on Jacobi elliptic functions \cite{NIST_Jacobi,Lawden}. With this compact formulation, we calculate the neoclassical susceptibility for arbitrary-wavelength axisymmetric electrostatic fluctuations in bounce-gyrokinetic theory \cite{Fong_Hahm,Brizard_2000}, and we recover the expression for the Rosenbluth-Hinton residual zonal flow in the long-wavelength limit \cite{ros98,Zonca} and improve on the work of Wang and Hahm \cite{Wang_Hahm}.

The remainder of the paper is organized as follows. In Sec.~\ref{sec:gc_orbits}, we briefly present the compact Jacobi-elliptic-function formulation of the trapped/passing-particle guiding-center orbits in simple axisymmetric tokamak geometry \cite{Brizard_2011,Brizard_2014}. Sec.~\ref{sec:bgy} provides a brief context for the use of orbit averages in bounce-gyrokinetic theory. In Sec.~\ref{sec:orbit}, we present the compact formulations of bounce and transit averaging associated with trapped/passing-particle guiding-center orbits in simple axisymmetric tokamak geometry. Formulas are given for arbitrary values of the pitch-angle parameter (which thus takes into account finite-orbit-width effects from the full range of trapped/passing orbits) as well as the short/long-wavelength limits. In Sec.~\ref{sec:nc_pol}, we present the calculation of the neoclassical susceptibility in the long-wavelength limit for axisymmetric electrostatic perturbations and recover the Rosenbluth-Hinton result.

\section{\label{sec:gc_orbits}Trapped/passing-particle Guiding-center Orbits}

The analytic representations of the trapped-particle and passing-particle guiding-center orbits in axisymmetric tokamak geometry were expressed in terms of Jacobi elliptic functions and integrals \cite{NIST_Jacobi,Lawden} in Ref.~\cite{Brizard_2011}. The trapped-particle and passing-particle guiding-center orbits are parameterized by the dimensionless pitch-angle parameter \cite{footnote}
\begin{equation}
\kappa({\cal E}, \mu, \psi) \;\equiv\; \frac{{\cal E} - \mu\,B_{\rm e}}{2\epsilon\;\mu B_{0}},
\label{eq:kappa}
\end{equation}
where ${\cal E}$ denotes the guiding-center's energy, $\mu$ denotes its guiding-center magnetic moment, and $\psi$ is the poloidal magnetic flux at the equatorial outside midplane. In addition, $\epsilon(\psi) < 1$ denotes the inverse aspect ratio and $B_{\rm e}(\psi) \equiv B_{0}\,(1 - \epsilon)$ denotes the equatorial magnetic-field strength on the outside midplane ($B_{0}$ denotes the magnetic-strength on the magnetic axis). 

For trapped-particle orbits $(\kappa < 1$), the parallel guiding-center momentum $p_{\|}$ and the poloidal angle $\vartheta$ are expressed in terms of Jacobi elliptic functions $({\rm cn}, {\rm sn})$ as \cite{Brizard_2011}:
\begin{eqnarray}
p_{\|} & = & p_{\|{\rm e}}\;{\rm cn}(\chi_{\rm b}|\kappa), \label{eq:p_cn} \\
\vartheta & = & 2\;\arcsin\left[ \sqrt{\kappa}\frac{}{}{\rm sn}(\chi_{\rm b}|\kappa)\right], \label{eq:theta_sn_b}
\end{eqnarray}
where the parallel momentum on the equatorial outside midplane is
\begin{equation}
p_{\|{\rm e}} \;=\; \sqrt{2m\,({\cal E} - \mu\,B_{\rm e})} \;\equiv\; 2\sqrt{\kappa}\;m\,\omega_{\|}\,R_{\|}, 
\label{eq:p||e_def}
\end{equation}
with the connection length $R_{\|} \equiv ds/d\vartheta$ defined as the rate of change of distance $s$ along a magnetic-field line as a function of the poloidal angle $\vartheta$ \cite{Brizard_2011} and $\omega_{\|}R_{\|} \equiv \sqrt{\epsilon\;\mu B_{0}/m}$. The bounce angle $\zeta_{\rm b} \equiv 
\chi_{\rm b}\,\nu_{\rm b}$ is defined in terms of the bounce factor
\begin{equation}
\nu_{\rm b}(\kappa) \;\equiv\; \frac{\pi}{2\,{\sf K}(\kappa)} \;=\; \frac{\omega_{\rm b}(\kappa)}{\omega_{\|}},
\label{eq:nu_b_def}
\end{equation}
where ${\sf K}(\kappa)$ denotes the complete elliptic integral of the first kind \cite{Lawden}. We note that $\nu_{\rm b}(\kappa)$ varies from 
$\nu_{\rm b}(0) = 1$ (deeply-trapped limit) to $\nu_{\rm b}(1) = 0$ (separatrix limit). 

For passing-particle orbits $(\kappa^{-1} < 1$), the parallel guiding-center momentum $p_{\|}$ and the poloidal angle $\vartheta$ are expressed in terms of Jacobi elliptic functions $({\rm dn}, {\rm sn})$  as \cite{Brizard_2011}:
\begin{eqnarray}
p_{\|} & = & p_{\|{\rm e}}\;{\rm dn}(\chi_{\rm t}|\kappa^{-1}), \label{eq:p_dn} \\
\vartheta & = & 2\;\arcsin\left[ \frac{}{}{\rm sn}(\chi_{\rm t}|\kappa^{-1})\right], \label{eq:theta_sn}
\end{eqnarray}
where the transit angle $\zeta_{\rm t} \equiv \chi_{\rm t}\,\nu_{\rm t}/\sqrt{\kappa}$ is defined in terms of the transit factor
\begin{equation}
\nu_{\rm t}(\kappa) \;\equiv\; \frac{\pi\,\sqrt{\kappa}}{{\sf K}(\kappa^{-1})} \;=\; \frac{\omega_{\rm t}(\kappa)}{\omega_{\|}},
\label{eq:nu_t_def}
\end{equation}
which varies from $\nu_{\rm t}(1) = 0$ (separatrix limit) to $\nu_{\rm t}(\kappa) \simeq 2\,\sqrt{\kappa}$ in the strongly-circulating limit 
($\kappa \gg 1$). 

\section{\label{sec:bgy}Bounce-gyrokinetic theory}

Bounce-gyrokinetic theory \cite{Fong_Hahm,Brizard_2000,Wang_Hahm} follows the same two-step Lie-transform approach of standard gyrokinetic theory 
\cite{Brizard_Hahm}. In the first step, the bounce-center phase-space transformation \cite{RGL_82, Cary_Brizard} decouples the fast bounce/transit motion of charged particles confined by a nonuniform magnetic field from the slow reduced drift motion in the absence of electromagnetic fluctuations. The bounce-center action is thus constructed as an adiabatic invariant and the unperturbed drift-motion dynamics is independent of the bounce-center angle. The introduction of low-frequency electromagnetic fluctuations destroys the invariance of the bounce-center action, which is restored with the help of a second (bounce-gyrocenter) phase-space transformation.

\subsection{Bounce-center transformation}

The bounce-center phase-space transformation from the reduced guiding-center coordinates to the bounce-center coordinates was recently 
\cite{Brizard_2014} solved explicitly in terms of the Jacobi elliptic functions for the case of axisymmetric tokamak geometry, where the magnetic field 
${\bf B} \equiv \nabla\xi\btimes\nabla\psi$ is expressed in terms of the poloidal magnetic flux $\psi$ and the Euler potential $\xi \equiv \varphi - 
q(\psi)\,\vartheta$. 

The transformation from the poloidal magnetic flux $\psi$ to the bounce-center magnetic-flux coordinate $\ov{\psi} \equiv \psi - \Delta\psi$ was expressed in terms of the poloidal-flux deviation $\Delta\psi$ from $\ov{\psi}$ \cite{Brizard_2011,Brizard_2014}:
\begin{eqnarray}
\Delta\psi & \equiv & (c/e)\,p_{\|}\,\ov{b}_{\varphi} \nonumber \\
 & = & \left\{ \begin{array}{l}
2\,\ov{B}_{\varphi}R_{\|}\,(\omega_{\|}/\Omega)\;\sqrt{\kappa}\;{\rm cn}(\chi_{\rm b}|\kappa) \\
 \\
2\,\ov{B}_{\varphi}R_{\|}\,(\omega_{\|}/\Omega)\;\sqrt{\kappa}\;{\rm dn}(\chi_{\rm t}|\kappa^{-1})
\end{array} \right.
\label{eq:Delta_psi}
\end{eqnarray}
where the parallel momentum $p_{\|}$ is either given by Eq.~\eqref{eq:p_cn} for trapped particles or Eq.~\eqref{eq:p_dn} for passing particles, and 
$\ov{B}_{\varphi} \equiv B_{\rm tor}\,R$ is the covariant toroidal component of the magnetic field evaluated at $\ov{\psi}$. 

The bounce-center transformation $\xi \rightarrow \ov{\xi} \equiv \xi - \Delta\xi$ is generated to first order by \cite{Brizard_2014}
\begin{equation}
\Delta\xi \;\equiv\; -\,\frac{c}{e} \left( \pd{S_{0}}{\psi} \;+\; p_{\|}\,b_{\varphi}\;\vartheta\,q^{\prime}\, \right),
\label{eq:G1_xi}
\end{equation}
where the scalar field $S_{0}(J,\zeta;\psi)$ generates the lowest-order canonical transformation $(p_{\|},s) \rightarrow (J,\zeta)$ from parallel guiding-center coordinates to bounce-center action-angle coordinates. Hence, the bounce-center Euler potential $\ov{\xi}$ is defined up to first order as
\begin{eqnarray}
\ov{\xi} & = & \left( \varphi \;+\; \frac{c}{e}\,\pd{S_{0}}{\psi} \right) \;-\; \vartheta \left[ q(\psi) \;-\frac{}{} \Delta\psi\;q^{\prime}(\psi)\right] \nonumber \\
& \equiv & \ov{\varphi} \;-\; \vartheta(\zeta,J)\;q(\ov{\psi}),
\label{eq:Xi_def}
\end{eqnarray}
which includes a transformation of the toroidal angle $\ov{\varphi} \equiv \varphi - \Delta\varphi$ generated by the canonical first-order component 
$\Delta\varphi = -\,(c/e)\,\partial S_{0}/\partial\psi$ and the safety factor $q(\ov{\psi})$ in Eq.~\eqref{eq:Xi_def} is now evaluated at the bounce-center magnetic-flux label $\ov{\psi}\equiv \psi - \Delta\psi$ defined by Eq.~\eqref{eq:Delta_psi}.

\subsection{Bounce-gyrocenter transformation}

We now consider electrostatic-field perturbations of the form \cite{Gang_Diamond}
\begin{equation}
\Phi_{1}(\psi)\,e^{-\,in\,\xi} \;\equiv\; \Phi_{1}(\ov{\psi} + \Delta\psi)\;e^{-\,in\,(\ov{\xi} + \Delta\xi)}, 
\label{eq:Phi1_def}
\end{equation}
where $n$ denotes the toroidal mode number and the right-hand side represents the perturbation $\Phi_{1}(\psi)\,\exp(-\,in\,\xi)$ expressed in terms of bounce-center coordinates. Hence, the perturbation \eqref{eq:Phi1_def} reintroduces a dependence on the poloidal angle $\vartheta$ (and therefore the bounce/transit angle $\zeta_{\rm b/t}$) through $(\Delta\psi,\Delta\xi)$, which thus destroys the invariance of the bounce/transit action $J_{\rm b/t}$. In the present paper, we focus our attention on axisymmetric perturbations $(n = 0)$, which play a crucial role in saturating the ion-temperature-gradient instability in axisymmetric tokamak plasmas \cite{Fong_Hahm,ros98,Lin},

One common approximation for an axisymmetric perturbation potential field $\Phi_{1}(\psi) \equiv \Phi_{1}(\ov{\psi} + \Delta\psi)$ is to assume that its radial dependence enters through an eikonal phase $\Theta(\psi)$ at the lowest order, i.e., $\Phi_{1}(\psi) \equiv \ov{\Phi}_{1}\,\exp[i\,\Theta(\psi)]$, where the eikonal amplitude $\ov{\Phi}_{1}$ and the radial wavevector ${\bf k}_{\bot} \equiv \nabla\Theta = k_{r}\,\nabla r$ are both weakly spatially dependent with respect to the orbit width. Hence, the dependence on the bounce/transit angle now appears through the eikonal-phase factor $\Theta(\psi) = \Theta(\ov{\psi}) + \Delta\Theta$, with $\Delta\Theta \equiv k_{r}\Delta\psi/(B_{\rm pol}\,R)$.

\section{\label{sec:orbit}Orbit Averaging in Bounce-gyrokinetic Theory}

In the present Section, we now look for explicit formulas for the orbit-averaged perturbation potential $\langle\Phi_{1}(\psi)\rangle_{\cal O} \equiv
\ov{\Phi}_{1}\;\langle\exp[i\,\Theta(\ov{\psi} + \Delta\psi)]\rangle_{\cal O}$. The Jacobi-elliptic representations \eqref{eq:p_cn}-\eqref{eq:theta_sn_b} and  \eqref{eq:p_dn}-\eqref{eq:theta_sn} of the guiding-center orbits can readily be applied to the orbit averages $({\cal O})$ of the eikonal-phase factor \cite{Wang_Hahm}
\begin{equation}
\langle e^{i\,\Theta(\psi)}\rangle_{\cal O} \;\simeq\; e^{i\Theta(\ov{\psi})}\;\left\langle e^{i\,\Delta\Theta}\right\rangle_{\cal O},
\label{eq:eikonal_ave}
\end{equation}
where 
\begin{equation}
\left\langle e^{i\,\Delta\Theta}\right\rangle_{\cal O} = \left\{ \begin{array}{l}
\left\langle\exp\left(i\,\alpha\frac{}{}\sqrt{\kappa}\,{\rm cn}(\chi_{\rm b}|\kappa)\right)\right\rangle_{\rm b} \\
 \\
\left\langle\exp\left(i\,\alpha\frac{}{}\sqrt{\kappa}\,{\rm dn}(\chi_{\rm t}|\kappa^{-1})\right)\right\rangle_{\rm t}
\end{array} \right.
\label{eq:eikonal_Jacobi}
\end{equation}
with
\begin{equation}
\alpha \;\equiv\; 2\;k_{r}R_{\|}\;\frac{B_{\rm tor}\,\omega_{\|}}{B_{\rm pol}\,\Omega} \;=\; \sqrt{2\epsilon}\;k_{r}\,\rho_{\rm pol},
\label{eq:alpha_def}
\end{equation}
defined in terms of the poloidal gyroradius $\rho_{\rm pol} \equiv \sqrt{2\mu\,B_{0}/m\Omega_{\rm pol}^{2}}$ and bounce/transit averaging 
$\langle\cdots\rangle_{\rm b/t}$ will be defined below.

\subsection{Work by Wang and Hahm}

Wang and Hahm \cite{Wang_Hahm} calculated the averaged eikonal-phase factor \eqref{eq:eikonal_ave} for the purpose of deriving the neoclassical polarization density according to bounce-center gyrokinetic theory. Two limits were considered: the short-wavelength limit $k_{r}\,\rho_{\rm pol} \gg 1$ (with $\alpha \gg 1$) and the long-wavelength limit $k_{r}\,\rho_{\rm pol} \ll 1$ (with $\alpha \ll 1$). 

Two additional limits were also considered when finite-orbit effects can be ignored: the deeply-trapped limit $(\kappa \ll 1$) for trapped-particle orbits and the strongly-circulating limit $(\kappa \gg 1$) for passing-particle orbits. It is in these limits that explicit analytic formulas for averages are typically presented in the literature (e.g., Ref.~\cite{Wang_Hahm}). As $\kappa \rightarrow 1$ (either from below, for trapped particles, or from above, for passing particles), however, the Jacobi-elliptic functions and integrals are required for an accurate description of significant finite-orbit effects, as will be shown below.

The most important difference between the work of Wang and Hahm \cite{Wang_Hahm} and the present work is our ability to deal explicitly with full finite-orbit effects with arbitrary pitch-angle parameter values: $0 \leq \kappa < 1$ for trapped particles (e.g., see Fig.~\ref{fig:bounce-coefficient}) and $\kappa > 1$ for passing particles (e.g., see Fig.~\ref{fig:transit_coefficient}).

\subsection{Bounce Average}

The bounce-averaging operation is formally defined as
\begin{equation}
\langle f\rangle_{\rm b} \;\equiv\; \frac{\omega_{\rm b}}{2\pi}\;\oint\;f\;\frac{ds}{v_{\|}},
\label{eq:bounce_def}
\end{equation}
where the integration cycle involves a round-trip between two turning points. By substituting 
\begin{equation} 
ds \;\equiv\; R_{\|}\;d\vartheta \;=\; R_{\|}\;\nu_{\rm b}\,(\partial\vartheta/\partial\zeta_{\rm b})\,d\chi_{\rm b} 
\label{eq:R||_def}
\end{equation}
and $v_{\|} = R_{\|}\omega_{\rm b}\,(\partial\vartheta/\partial\zeta_{\rm b})$, we find $ds/v_{\|} = d\chi_{\rm b}/\omega_{\|}$. Since $\omega_{\rm b}/
2\pi = \omega_{\|}/(4\,{\sf K})$, the bounce-averaged eikonal factor \eqref{eq:eikonal_ave} becomes
\begin{eqnarray}
\left\langle e^{i\Delta\Theta}\right\rangle_{\rm b} & = & \left\langle \exp\left[i\alpha\;\sqrt{\kappa}\frac{}{}
{\rm cn}(\chi_{\rm b}|\kappa)\right] \right\rangle_{\rm b} 
\label{eq:eikonal_bounce_ave} \\
 & = & \left(\int_{-2{\sf K}}^{2{\sf K}}\; \exp\left[i\alpha\;\sqrt{\kappa}\frac{}{}{\rm cn}(\chi_{\rm b}|\kappa)\right] \frac{d\chi_{\rm b}}{4{\sf K}} \right).
\nonumber
\end{eqnarray}
This average can be expressed in terms of the multi-variable generalized Bessel functions of Dattoli et al.~\cite{dat92,dat98}.  In the present work, we explicitly express these generalized functions in terms of products of standard Bessel functions.

We note that in the deeply-trapped limit $(\kappa \rightarrow 0)$ and the barely-trapped limit $(\kappa \rightarrow 1)$, the bounce-averaged eikonal factor \eqref{eq:eikonal_bounce_ave} becomes
\begin{equation}
\lim_{\kappa \rightarrow 0}\;\left\langle e^{i\Delta\Theta}\right\rangle_{\rm b} \;=\; 1 \;=\; \lim_{\kappa \rightarrow 1}\;\langle 
e^{i\Delta\Theta}\rangle_{\rm b},
\label{eq:bounce_limit}
\end{equation}
where we used the limits
\begin{equation}
\left. \begin{array}{rcl}
\lim_{\kappa \rightarrow 0}\sqrt{\kappa}\,{\rm cn}(\chi_{\rm b}|\kappa) & = & 0 \\
 &  & \\
\lim_{\kappa \rightarrow 1}\sqrt{\kappa}\,{\rm cn}(\chi_{\rm b}|\kappa) & = & \lim_{z \rightarrow \infty}{\rm sech}\,z = 0
\end{array} \right\}.
\end{equation}

\begin{figure}
\epsfysize=2in
\epsfbox{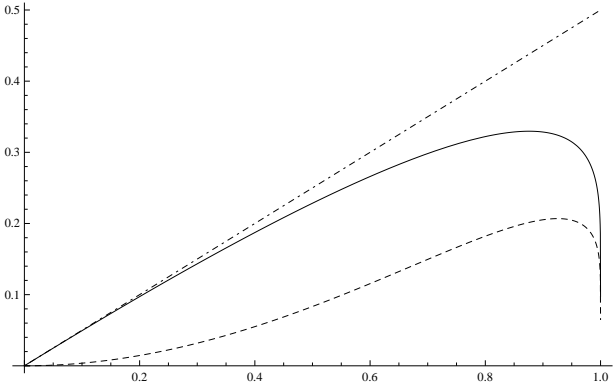}
\caption{Plots of $B_{1}$ (solid) and $B_{2}$ (dashed) in the range $0 \leq \kappa < 1$, with the approximation $B_{1}(\kappa) \simeq \kappa/2$ (shown as a dot-dashed line) in the deeply-trapped limit $\kappa \ll 1$.}
\label{fig:Bounce_long}
\end{figure}

The Taylor expansion of Eq.~\eqref{eq:eikonal_bounce_ave} in powers of $\alpha$ yields
\begin{eqnarray}
\left\langle e^{i\Delta\Theta}\right\rangle_{\rm b} & = & \sum_{n = 0}^{\infty} \frac{(i\alpha\sqrt{\kappa})^{n}}{n!}\;\int_{-2{\sf K}}^{2{\sf K}}\; 
{\rm cn}^{n}(\chi_{\rm b}|\kappa)\;\frac{d\chi_{\rm b}}{4{\sf K}} \nonumber \\
 & = & \sum_{m = 0}^{\infty} \frac{(-\alpha^{2}\kappa)^{m}}{(2m)!}\;\int_{-2{\sf K}}^{2{\sf K}}\; {\rm cn}^{2m}(\chi_{\rm b}|\kappa)\;
\frac{d\chi_{\rm b}}{4{\sf K}} \nonumber \\
 & \equiv & \sum_{m = 0}^{\infty} \frac{(-1)^{m}\;\alpha^{2m}}{(2m)!}\;B_{m}(\kappa),
\label{eq:Bounce_Taylor}
\end{eqnarray}
where the first integral vanishes if $n$ is odd and the coefficients
\begin{equation}
B_{m}(\kappa) \;\equiv\; \kappa^{m}\,\int_{-2{\sf K}}^{2{\sf K}}{\rm cn}^{2m}(\chi_{\rm b}|\kappa)\,\frac{d\chi_{\rm b}}{4{\sf K}} 
\label{eq:B_def}
\end{equation}
satisfy the recurrence relation for $m \geq 1$ (adapted from exercise 3.15 of Ref.~\cite{Lawden}):
\begin{eqnarray}
B_{m+1} & = & \frac{2m}{(2m+1)}\,(2\kappa-1)\,B_{m} \nonumber \\
 &  &+\; \frac{(2m-1)}{(2m+1)}\,(1 - \kappa)\kappa\,B_{m-1},
\end{eqnarray}
with $B_{0} = 1$ and
\begin{equation}
\left. \begin{array}{rcl}
B_{1} & = & {\sf E}/{\sf K} \;-\; (1 - \kappa) \\
 &  & \\
B_{2} & = & \frac{2}{3}\,(2\kappa - 1)\,{\sf E}/{\sf K} \;-\; (1 - \kappa)\,\left(\kappa - \frac{2}{3}\right)
\end{array} \right\}.
\label{eq:B12_def}
\end{equation}
Hence, in the long-wavelength limit $(\alpha \ll 1)$, we find from Eq.~\eqref{eq:Bounce_Taylor}
\begin{equation}
\left\langle e^{i\Delta\Theta}\right\rangle_{\rm b} \;\simeq\; 1 \;-\; \frac{1}{2}\;\alpha^{2}\,B_{1}(\kappa),
\label{eq:Bounce_long}
\end{equation}
where $B_{1}(\kappa)$ and $B_{2}(\kappa)$ are shown in Fig.~\ref{fig:Bounce_long}, with the approximation $B_{1}(\kappa) \simeq \kappa/2$ in the deeply-trapped limit $(\kappa \ll 1)$ also shown.

\subsubsection{Fourier expansion}

\begin{figure}
\epsfysize=2in
\epsfbox{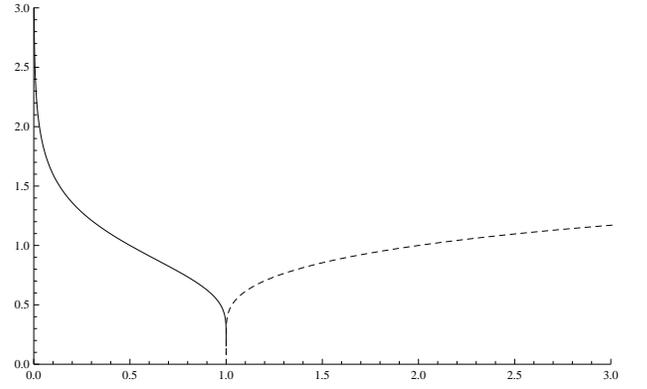}
\caption{Plot of $|\tau(\kappa)| \equiv {\sf K}(1 - \kappa)/{\sf K}(\kappa)$ (solid) and $|\ov{\tau}(\kappa)| \equiv {\sf K}(1 - \kappa^{-1})/
{\sf K}(\kappa^{-1})$ (dashed) in the range $0 \leq \kappa < 1$ and $\kappa > 1$, respectively. Limiting values are $|\tau(\kappa)| \rightarrow \infty$ as $\kappa \rightarrow 0$ and $|\tau(1)| \equiv 0 \equiv |\ov{\tau}(1)|$.}
\label{fig:tau}
\end{figure}

\begin{figure}
\epsfysize=2in
\epsfbox{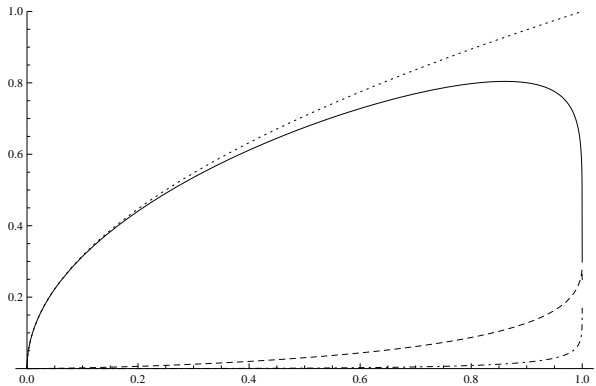}
\caption{Plots of $a_{1}$ (solid), $a_{3}$ (dashed), and $a_{5}$ (dot-dashed) in the range $0 \leq \kappa < 1$. The dotted curve shows that $a_{1}(\kappa)
\simeq \sqrt{\kappa}$ in the deeply-trapped limit $\kappa \ll 1$.}
\label{fig:bounce-coefficient}
\end{figure}

The short-wavelength limit $(\alpha \gg 1)$ requires a different approach than the Taylor expansion \eqref{eq:Bounce_Taylor}. In order to explicitly evaluate the bounce-angle averaging in Eq.~\eqref{eq:eikonal_bounce_ave}, we now introduce the Fourier series (formula 8.7.7 of Ref.~\cite{Lawden})
\begin{equation}
\sqrt{\kappa}\;{\rm cn}(\zeta_{\rm b}/\nu_{\rm b}|\kappa) \;=\; 2\,\nu_{\rm b}\;\sum_{n=0}^{\infty}\;\frac{\cos[(2 n + 1)\,\zeta_{\rm b}]}{
\cos[(n + 1/2)\,\pi\,\tau]},
\end{equation}
where the Jacobi parameter $\tau(\kappa) \equiv i\,{\sf K}(1 - \kappa)/{\sf K}(\kappa)$ is shown in Fig.~\ref{fig:tau} (solid curve). The bounce-averaged eikonal-phase factor \eqref{eq:eikonal_bounce_ave} then becomes
\begin{eqnarray}
\left\langle e^{i\Delta\Theta}\right\rangle_{\rm b} & = & \left\langle \prod_{n = 0}^{\infty}e^{(i\alpha\,a_{2n+1}\cos[(2n+1)\zeta_{\rm b}])}\right\rangle_{\rm b},
\label{eq:eikonal_ave_final}
\end{eqnarray}
where the bounce-averaging coefficients are defined as
\begin{equation}
a_{2n+1}(\kappa) \;\equiv\; \frac{2\;\nu_{\rm b}(\kappa)}{\cos[(n + 1/2)\;\pi\,\tau(\kappa)]}.
\label{eq:a_def}
\end{equation} 
Figure \ref{fig:bounce-coefficient} shows the bounce-averaging coefficients \eqref{eq:a_def} in the range $0 \leq \kappa < 1$ for $n = 0$ (solid), 
$n = 1$ (dashed), and $n = 2$ (dot-dashed). We note that $a_{2n+1}(\kappa) \rightarrow 0$ as $\kappa \rightarrow 0$ and 1 for all $n$, so that 
Eq.~\eqref{eq:bounce_limit} is satisfied. We also note that $a_{1}$ dominates in the deeply-trapped limit $(\kappa \ll 1$), where the approximation 
$a_{1} \simeq \sqrt{\kappa}$ is used by Wang and Hahm \cite{Wang_Hahm}.

By using the Fourier-Bessel series
\[ \exp\left(i\,\alpha\,a_{k}\frac{}{}\cos(k\zeta_{\rm b})\right) \;=\; \sum_{m = -\infty}^{\infty}\;i^{m}\;e^{im\,k\zeta_{\rm b}}\; 
J_{m}(\alpha\,a_{k}), \]
we obtain the bounce-averaged expression in Eq.~\eqref{eq:eikonal_ave_final}:
\begin{eqnarray}
\left\langle e^{i\Delta\Theta}\right\rangle_{\rm b} & = & {\cal J}_{0}(\alpha,\kappa) + \sum_{{\bf m}\neq 0}\;{\cal J}_{\bf m}(\alpha, \kappa),
\label{eq:J_0m}
\end{eqnarray}
where the ``fundamental'' contribution is defined by the infinite Bessel product
\begin{eqnarray}
{\cal J}_{0}(\alpha,\kappa) & \equiv & \prod_{n = 0}^{\infty}J_{0}(\alpha\,a_{2n+1}) \nonumber \\
 & = & J_{0}(\alpha\,a_{1})\,J_{0}(\alpha\,a_{3})\;\cdots,
\label{eq:J0_def}
\end{eqnarray}
and the ``harmonic'' contribution is defined by the infinite Bessel product
\begin{equation}
{\cal J}_{\bf m}(\alpha,\kappa) \;\equiv\; 2\,i^{(m_{0}+m_{1}+\cdots)}\;J_{m_{0}}(\alpha\,a_{1})\;J_{m_{1}}(\alpha\,a_{3})\;\cdots,
\label{eq:Jm_def}
\end{equation}
with the constraint on the infinite-dimensional integer-vector ${\bf m} = (m_{0}, m_{1}, m_{2}, \cdots)$:
\[ 0 \;\equiv\; \sum_{n = 0}^{\infty}\;(2n+1)\,m_{n} \;=\; m_{0} + 3\,m_{1} + 5\,m_{2} + \cdots. \]
We note that, depending on the relative sizes of the bounce-averaging coefficients \eqref{eq:a_def}, the fundamental \eqref{eq:J0_def} and harmonic 
\eqref{eq:Jm_def} contributions to the bounce average \eqref{eq:J_0m} may include a large number of terms, especially as we approach the trapped-passing boundary.

Figure \ref{fig:bounce-coefficient} shows that for trapped-particle orbits that are far from the trapped-passing boundary $(\kappa < 0.8)$, only the coefficients $a_{1}$ (solid line) and $a_{3}$ (dashed line) are large enough to contribute to the bounce-average \eqref{eq:eikonal_ave_final}. Hence, the fundamental contribution \eqref{eq:J0_def} can be approximated as
\begin{equation}
{\cal J}_{0}(\alpha,\kappa) \;\simeq\; J_{0}(\alpha\,a_{1})\;J_{0}(\alpha\,a_{3}),
\end{equation}
while the harmonic contribution \eqref{eq:Jm_def} can be approximated as
\begin{equation}
{\cal J}_{M}(\alpha,\kappa) \;\equiv\; \sum_{m = 1}^{M}\;  2\,(-1)^{m}\,J_{3m}(\alpha\,a_{1})\,J_{m}(\alpha\,a_{3}),
\label{eq:JM_def}
\end{equation}
where, for practical applications, we truncate the harmonic contribution at a finite order $M$.

\subsubsection{Limiting cases}

\begin{figure}
{\includegraphics[width=0.45\textwidth]{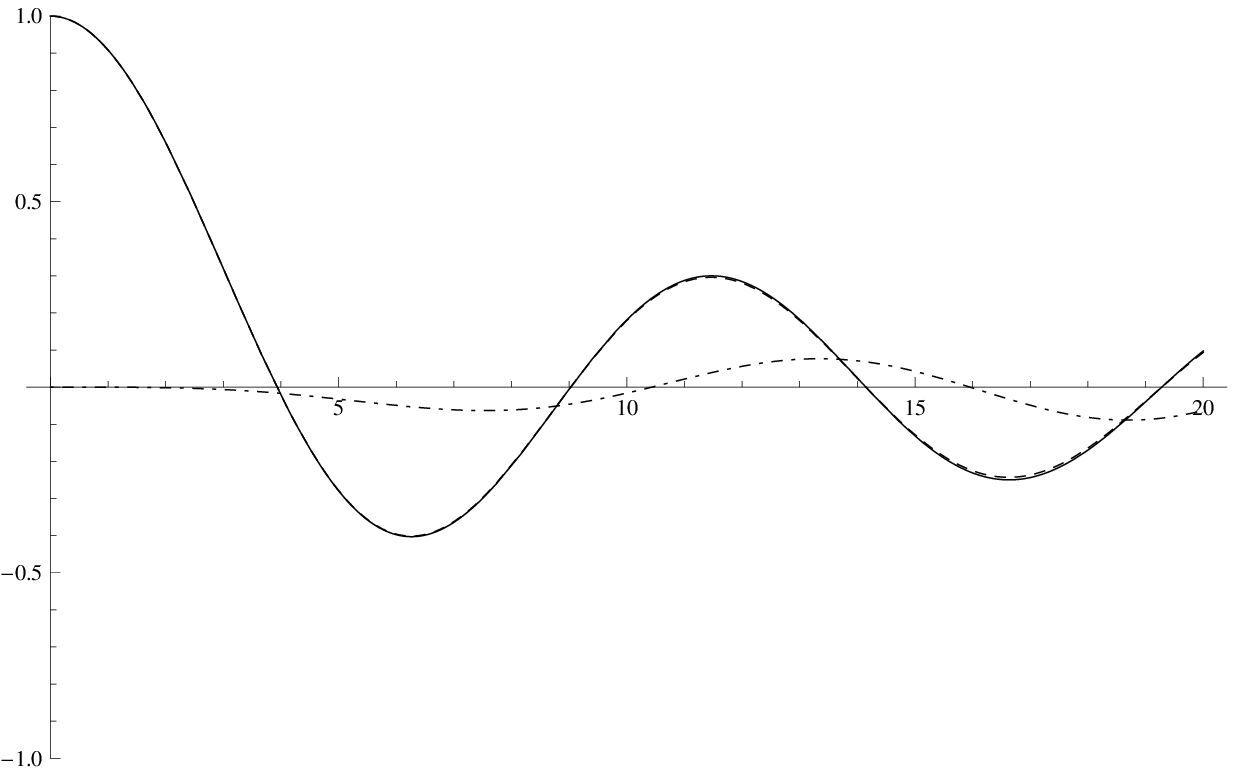}}
{\includegraphics[width=0.45\textwidth]{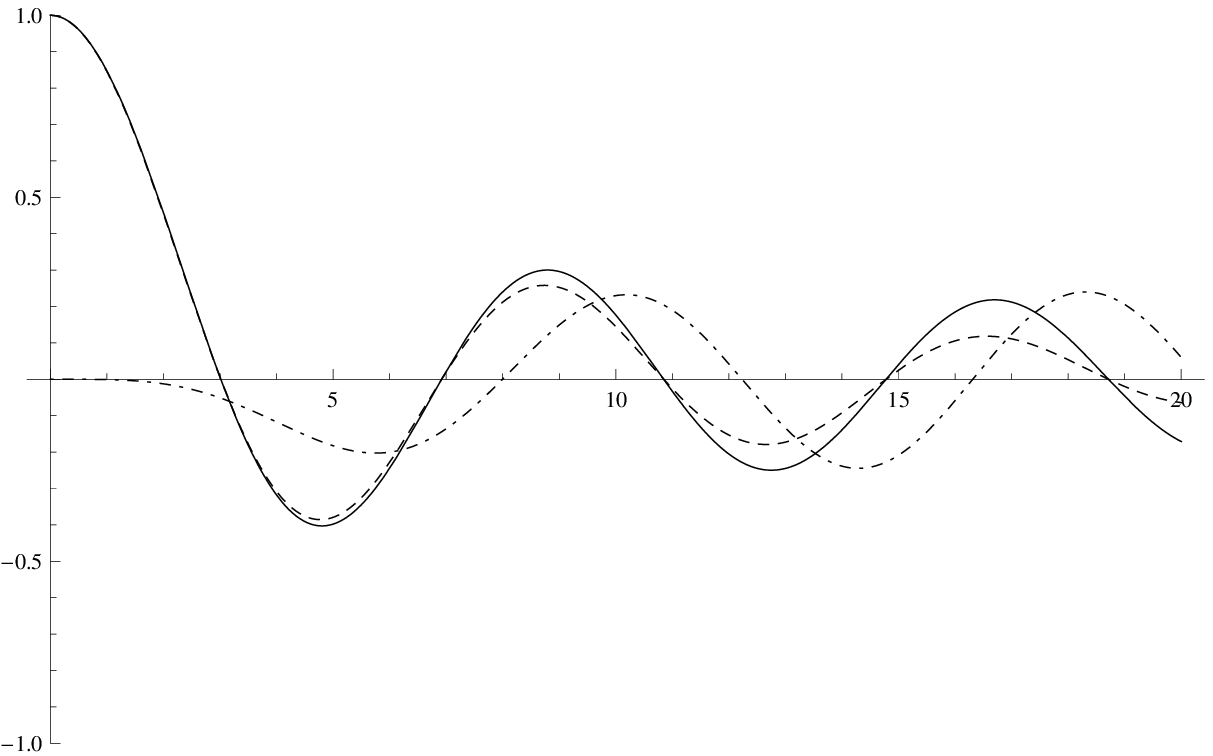}}
\caption{Plots of $J_{0}(\alpha\,a_{1})$ (solid), ${\cal J}_{0}(\alpha,\kappa) \simeq J_{0}(\alpha\,a_{1})\;J_{0}(\alpha\,a_{3})$ (dashed), and 
${\cal J}_{M=1}(\alpha,\kappa) \simeq -2\,J_{3}(\alpha\,a_{1})\,J_{1}(\alpha\,a_{3})$ (dot-dashed) in the range $0 \leq \alpha \leq 20$ for $\kappa = 0.4$ (top) and $\kappa = 0.8$ (bottom).}
\label{fig:Bounce_average}
\end{figure}

In the deeply-trapped limit $(\kappa \ll 1)$, the fundamental contribution \eqref{eq:J0_def} agrees very well with the deeply-trapped limit: 
$J_{0}(\alpha\,a_{1}) \simeq J_{0}(\alpha\,\sqrt{\kappa})$ used by Wang and Hahm \cite{Wang_Hahm}. In the moderately-trapped regime $(0.4 \leq 
\kappa \leq 0.8)$, however, Fig.~\ref{fig:Bounce_average} (top: $\kappa = 0.4$; bottom: $\kappa = 0.8$) shows that, in the short-wavelength limit 
$\alpha \gg 1$, we begin to see significant departures from $J_{0}(\alpha\,a_{1})$ as $\kappa$ increases and a progressive influence of the harmonic contributions ${\cal J}_{M=1}(\alpha, \kappa) \simeq -2\,J_{3}(\alpha\,a_{1})\,J_{1}(\alpha\,a_{3})$ over the fundamental contribution ${\cal J}_{0}(\alpha,\kappa) \simeq J_{0}(\alpha\,a_{1})\,J_{0}(\alpha\,a_{3})$. We note, however, that for $\alpha < 2$, we can reliably use 
\begin{equation}
\langle e^{i\Delta\Theta}\rangle_{\rm b} \;\simeq\; J_{0}(\alpha\,a_{1}) 
\label{eq:bounce_approx}
\end{equation}
for pitch-angle-parameter values almost up to $\kappa = 1$.

Lastly, we note that the long-wavelength limit of the Fourier-Bessel formula \eqref{eq:J_0m} yields the expansion
\begin{equation}
\langle e^{i\Delta\Theta}\rangle_{\rm b} \;=\; 1 \;-\; \frac{\alpha^{2}}{4}\;\left( a_{1}^{2}(\kappa) \;+\; a_{3}^{2}(\kappa) \;+\frac{}{} \cdots
\right),
\label{eq:long_bounce_Bessel}
\end{equation}
which agrees with the Taylor-expansion \eqref{eq:Bounce_long} (see Fig.~\ref{fig:nc_trapped}).

\subsection{Transit Average}

For passing-particle guiding-center orbits $(\kappa > 1)$, the transit-averaged eikonal-phase factor is defined as
\begin{eqnarray}
\langle e^{i\Delta\Theta}\rangle_{\rm t} & = & \left\langle \exp\left[i\sigma\,\alpha\;\sqrt{\kappa}\frac{}{}
{\rm dn}(\chi_{\rm t}|\kappa^{-1})\right] \right\rangle_{\rm t} 
\label{eq:eikonal_transit_ave} \\
 & = & \left( \int_{0}^{2\ov{\sf K}}\exp\left[i\sigma\,\alpha\sqrt{\kappa}\frac{}{}{\rm dn}(\chi_{\rm t}|\kappa^{-1})\right]
\frac{d\chi_{\rm t}}{2\ov{\sf K}} \right),
\nonumber
\end{eqnarray}
where $\ov{\sf K} \equiv {\sf K}(\kappa^{-1})$, and $\sigma$ denotes the sign of $p_{\|}$: $\sigma = +1$ for co-passing particles; and $\sigma = -1$ for counter-passing particles. Again, this average can be expressed in terms of the multi-variable generalized Bessel functions of Dattoli et al.~\cite{dat92,dat98}.

In the limit where we are approaching the trapped-passing boundary $\kappa = 1$, we find
\begin{equation}
\lim_{\kappa \rightarrow 1}\;\left\langle e^{i\Delta\Theta}\right\rangle_{\rm t} \;=\; 1,
\label{eq:transit_limit}
\end{equation}
where we used $\lim_{\kappa \rightarrow 1}\;\sqrt{\kappa}\;{\rm dn}(\chi_{\rm t}|\kappa^{-1}) = 0$.

\begin{figure}
\epsfysize=2in
\epsfbox{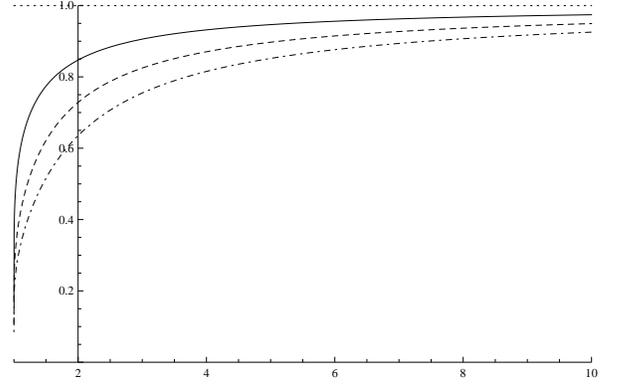}
\caption{Plots of $T_{1}$ (solid), $T_{2}$ (dashed), and $T_{3}$ (dot-dashed) in the range $1 < \kappa \leq 10$.}
\label{fig:transit_T_coefficient}
\end{figure}

The Taylor expansion of Eq.~\eqref{eq:eikonal_transit_ave} in powers of $\alpha$ yields
\begin{eqnarray}
\left\langle e^{i\Delta\Theta}\right\rangle_{\rm t} & = & \sum_{n=0}^{\infty} \frac{(i\sigma\alpha\sqrt{\kappa})^{n}}{n!}\;\int_{0}^{2\ov{\sf K}}
{\rm dn}^{n}(\chi_{\rm t}|\kappa^{-1})\;\frac{d\chi_{\rm t}}{2\ov{\sf K}} \nonumber \\
 & \equiv & \sum_{n=0}^{\infty} \frac{(i\sigma\alpha\sqrt{\kappa})^{n}}{n!}\;T_{n}(\kappa),
\label{eq:Transit_Taylor}
\end{eqnarray}
where the coefficients 
\begin{equation}
T_{n}(\kappa) \;\equiv\; \int_{0}^{2\ov{\sf K}}{\rm dn}^{n}(\chi_{\rm t}|\kappa^{-1})\,\frac{d\chi_{\rm t}}{2\ov{\sf K}} 
\label{eq:T_def}
\end{equation}
satisfy the recurrence relation for $n \geq 2$ (adapted from exercise 3.15 of Ref.~\cite{Lawden}):
\begin{eqnarray}
T_{n+2} & = & \frac{n}{(n+1)}\,(2 - \kappa^{-1})\,T_{n} \nonumber \\
 &  &-\; \frac{(n-1)}{(n+1)}\,(1 - \kappa^{-1})\,T_{n-2},
\end{eqnarray}
with $T_{0} = 1$ and
\begin{equation}
\left. \begin{array}{rcl}
T_{1} & = & \pi/2\ov{\sf K} \\
 &  & \\
T_{2} & = & \ov{\sf E}/\ov{\sf K} \\
 &  & \\
T_{3} & = & (2 - \kappa^{-1})\,\pi/4\ov{\sf K}
\end{array} \right\}.
\label{eq:T_123_def}
\end{equation}
Hence, in the long-wavelength limit $(\alpha \ll 1)$, we find from Eq.~\eqref{eq:Transit_Taylor}
\begin{equation}
\left\langle e^{i\Delta\Theta}\right\rangle_{\rm t} \;\simeq\; 1 \;+\; i\sigma\;\alpha\sqrt{\kappa}\;T_{1}(\kappa) \;-\; \frac{1}{2}\,\alpha^{2}\,\kappa\;
T_{2}(\kappa).
\label{eq:Transit_long}
\end{equation}
Since $T_{n}(\kappa) < 1$ for all $n$ (see Fig.~\ref{fig:transit_T_coefficient}), we note that $\langle e^{i\Delta\Theta}\rangle_{\rm t} < \exp(i\sigma\,
\alpha\sqrt{\kappa})$.

\subsubsection{Fourier expansion}

\begin{figure}
\epsfysize=2in
\epsfbox{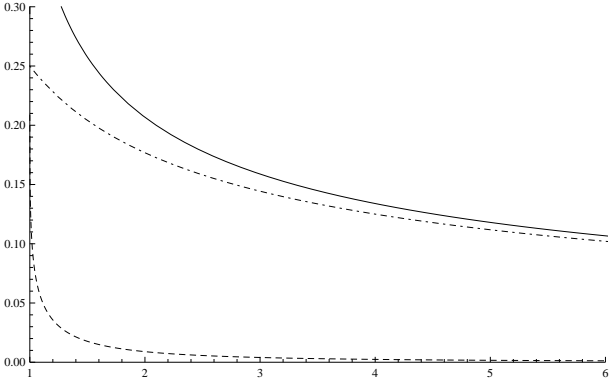}
\caption{Plots of $b_{2}$ (solid) and $b_{4}$ (dashed) in the range $1 < \kappa \leq 6$. In the strongly-circulating limit $(\kappa \gg 1$), we find
$b_{2} \simeq 1/(4\,\sqrt{\kappa})$ (dot-dashed).}
\label{fig:transit_coefficient}
\end{figure}

The short-wavelength limit $(\alpha \gg 1)$ requires a different approach than the Taylor expansion \eqref{eq:Transit_Taylor}. We thus introduce the Fourier series (formula 8.7.8 of Ref.~\cite{Lawden})
\begin{equation}
\sqrt{\kappa}\,{\rm dn}(\chi_{\rm t}|\kappa^{-1}) \;=\; \frac{\nu_{\rm t}}{2} \;+\; \nu_{\rm t}\;\sum_{n=1}^{\infty}\;
\frac{\cos(2n\,\zeta_{\rm t})}{\cos(n\pi\,\ov{\tau})},
\end{equation}
where $\ov{\tau}(\kappa) \equiv i\,{\sf K}(1 - \kappa^{-1})/{\sf K}(\kappa^{-1})$ is shown in Fig.~\ref{fig:tau} (dashed curve). The transit-averaged eikonal-phase factor \eqref{eq:eikonal_transit_ave} therefore becomes
\begin{eqnarray}
\left\langle e^{i\Delta\Theta}\right\rangle_{\rm t} & = & e^{i\sigma\alpha\nu_{\rm t}/2}\left\langle 
\prod_{n = 1}^{\infty}e^{(i\sigma\alpha b_{2n}\cos(2n\zeta_{\rm t}))}\right\rangle_{\rm t},
\label{eq:eikonal_transit_final}
\end{eqnarray}
where the transit-averaging coefficients are defined as
\begin{equation}
b_{2n}(\kappa) \;\equiv\; \frac{\nu_{\rm t}(\kappa)}{\cos(n\pi\,\ov{\tau})}.
\label{eq:b_def}
\end{equation}
Figure \ref{fig:transit_coefficient} shows the transit-averaging coefficients $b_{2}$ (solid) and $b_{4}$ (dashed). We note that the transit-averaging coefficient $b_{2}$ is approximated as $1/4\sqrt{\kappa}$ (dot-dashed curve in Fig.~\ref{fig:transit_coefficient}) in the strongly-circulating limit
 $(\kappa \gg 1)$.

Next, we use the Fourier-Bessel series
\[ \exp\left(i\,\sigma\alpha\,b_{k}\frac{}{}\cos(k\zeta_{\rm t})\right) \;=\; \sum_{m = -\infty}^{\infty}\;(i\sigma)^{m}\,e^{im\,k\zeta_{\rm t}}\; 
J_{m}(\alpha\,b_{k}), \]
to find that, for passing-particle guiding-center orbits far from the trapped-passing boundary (i.e., $\kappa > 2$), the transit average 
\eqref{eq:eikonal_transit_final} can be approximated as
\begin{eqnarray}
\left\langle e^{i\Delta\Theta}\right\rangle_{\rm t} & \simeq & e^{i\sigma\,\alpha\,\nu_{\rm t}/2}\,J_{0}(\alpha\,b_{2}). 
\label{eq:transit_approx}
\end{eqnarray}
Lastly, as was the case with the bounce-averaging result, the number of transit-averaging coefficients $b_{2n}$ needed to evaluate the transit average 
\eqref{eq:eikonal_transit_final} increases as we approach the trapped-passing boundary $(\kappa \rightarrow 1)$.

\subsubsection{Limiting cases}

In the short-wavelength $(\alpha \gg 1)$ and strongly-circulating $(\kappa \gg 1)$ limits, the transit average 
\eqref{eq:transit_approx} is approximated as
\begin{eqnarray}
\left\langle e^{i\Delta\Theta}\right\rangle_{\rm t} & \simeq & \left(\frac{2\,\sqrt{\kappa}}{\pi\,\alpha}\right)^{1/2}\exp\left[i\left(\sigma\,\alpha\,\sqrt{\kappa} \pm \frac{\pi}{4}\right) \right],
\label{eq:Transit_strong}
\end{eqnarray}
which is consistent with the stationary-phase result (45) of Wang and Hahm \cite{Wang_Hahm}, with
\[ J_{0}(\alpha\,b_{2}) \;\simeq\; \sqrt{\frac{1}{2\pi\,\alpha\,b_{2}}}\;\left[ e^{i(\alpha\,b_{2} - \pi/4)} \;+\; e^{-i(\alpha\,b_{2} - \pi/4)}  
\right], \]
with $\nu_{\rm t} \simeq 2\,\sqrt{\kappa}$ and $b_{2} \simeq 1/4\sqrt{\kappa} \ll \sqrt{\kappa}$ in the strongly-circulating limit $\kappa \gg 1$. 

In the long-wavelength limit $(\alpha \ll 1)$, we find
\begin{eqnarray}
e^{i\sigma\,\alpha\,\nu_{\rm t}/2}\,J_{0}(\alpha\,b_{2}) & \simeq & e^{i\sigma\,\alpha\,\nu_{\rm t}/2} \left( 1 \;-\; \frac{\alpha^{2}}{4}\,
b_{2}^{2}(\kappa) + \cdots \right) \nonumber \\
 & = & 1 \;+\; i\sigma\,\alpha\;\frac{\nu_{\rm t}}{2} \;-\; \frac{\alpha^{2}}{8}\;\nu_{\rm t}^{2} \;+\; \cdots \nonumber \\
 & = & 1 \;+\; i\sigma\,\frac{2\alpha}{\pi}\,\sqrt{\kappa}\;{\sf E}(\kappa^{-1}) \nonumber \\
 &  &-\; \frac{\alpha^{2}}{2} \left( \kappa - \frac{1}{2} \right) \;+\; \cdots,
\end{eqnarray}
where we used the identity $1/{\sf K}(\kappa^{-1}) \equiv (2/\pi)^{2}\,{\sf E}(\kappa^{-1})$ and the expansion $\nu_{\rm t}^{2}/2 \simeq 2\,\kappa - 1 + \cdots \gg b_{2}^{2}(\kappa)$ valid for $\kappa \gg 1$. This result agrees with Eq.~\eqref{eq:Transit_long} and is identical to Eq.~(48) of Wang and Hahm \cite{Wang_Hahm}.

\section{\label{sec:nc_pol}Neoclassical polarization in the Long-Wavelength Limit}

In this Section, we consider an application of the compact formulation of orbit averaging presented in Sec.~\ref{sec:orbit} by deriving the formula for the neoclassical polarization in the long-wavelength limit $(\alpha \ll 1$).

One direct consequence of dynamical reduction associated with the phase-space transformation ${\cal T}_{\epsilon}: {\bf z} \rightarrow {\bf Z} \equiv 
{\cal T}_{\epsilon}{\bf z}$ is the introduction of polarization effects in the reduced Maxwell equations \cite{Brizard_reduced_1,Brizard_reduced_2,Brizard_reduced_3}. The reduced polarization charge density $\Delta\varrho \equiv \varrho - \ov{\varrho}$, defined as the difference between the particle charge density $\varrho$ and the reduced charge density $\ov{\varrho}$, can be expressed in terms of the general expression
\begin{eqnarray}
\Delta\varrho & \equiv & \sum e\int \left( \left\langle{\sf T}_{\epsilon}F\right\rangle_{\cal O} \;-\frac{}{} F\right)\,d^{3}P,
\label{eq:pol_def}
\end{eqnarray}
where $\langle\cdots\rangle_{\cal O}$ denotes orbit-averaging with respect to the fast dynamics and the particle Vlasov distribution $f \equiv {\sf T}_{\epsilon}F$ is expressed in terms of the reduced pull-back operator ${\sf T}_{\epsilon}$ acting on the reduced Vlasov distribution $F$.

In bounce-gyrokinetic theory \cite{Wang_Hahm,Fong_Hahm,Brizard_2000}, the reduced polarization charge density naturally divides into classical (cl) and neoclassical (nc) contributions: $\Delta\varrho \equiv \Delta\varrho_{\rm cl} + \Delta\varrho_{\rm nc}$. The classical polarization contributions are associated with the gyromotion dynamical reduction carried out through the guiding-center (gc) and gyrocenter (gy) phase-space transformations, while the neoclassical polarization contributions are associated with the bounce-motion dynamical reduction carried out through the bounce-center (bc) and bounce-gyrocenter (bgy) phase-space transformations. 

We note that the classical and neoclassical contributions due to the guiding-center and bounce-center transformations explicitly depend on the background magnetic-field nonuniformity \cite{Brizard_reduced_3}. These contributions, which are omitted here, are normally implicitly included in the definitions of the gyrocenter and bounce-gyrocenter densities (e.g., see Eq.~(24) of 
Ref.~\cite{Wang_Hahm}). The classical and neoclassical contributions due to the gyrocenter and bounce-gyrocenter transformations, on the other hand, explicitly depend on the fluctuating electrostatic potential $\Phi_{1}$, and these contributions are now considered separately. 

First, the classical gyrocenter polarization charge density is expressed as
\begin{eqnarray}
\Delta\varrho_{\rm cl} & \equiv & -\sum \int \frac{e^{2}F}{T}\left\langle {\sf T}_{\rm gc}\left(\Phi_{1{\rm gc}} -\frac{}{} \langle\Phi_{1{\rm gc}}
\rangle_{\rm g}\right)\right\rangle_{\rm g} d^{3}P \nonumber \\
 & = & -\sum \frac{e^{2}\ov{\Phi}_{1}}{T}\,e^{i\Theta(\psi)}\int \left(1 -\frac{}{} J_{0}^{2}(k_{r}\rho)\right)\;F\,d^{3}P \nonumber \\
 & \equiv & -\sum \frac{e^{2}\,N}{T}\,\ov{\Phi}_{1}e^{i\Theta(\psi)}\;\chi_{\rm cl}, 
\label{eq:classical} 
\end{eqnarray}
where the gyroangle-averaged eikonal factor $\langle \exp(i{\bf k}\bdot\vb{\rho})\rangle_{\rm g} \equiv J_{0}(k_{r}\rho)$ is expressed in terms of the zeroth-order Bessel function and we assumed that the bounce-gyrocenter distribution $F \equiv N\,\exp(-{\cal E}/T)/(2\pi\,mT)^{3/2}$ is a Maxwellian distribution. Here, the classical gyrocenter susceptibility is defined in the long-wavelength limit $(k_{r}\rho_{\rm T} \ll 1)$ as
\begin{equation}
\chi_{\rm cl} \;\equiv\; 1 - I_{0}\left[(k_{r}\rho_{\rm T})^{2}\right]\,\exp\left[-(k_{r}\rho_{\rm T})^{2}\right] \;\simeq\; (k_{r}\rho_{\rm T})^{2},
\label{eq:chi_cl}
\end{equation}
where $I_{0}\left[(k_{r}\rho_{\rm T})^{2}\right]$ denotes the zeroth-order modified Bessel function, with the dimensionless parameter $k_{r}\rho_{\rm T}$ defined in terms of the thermal gyroradius $\rho_{\rm T} = (T/m\Omega^{2})^{1/2}$. Because of the large ion-electron mass ratio $m_{\rm i}/m_{\rm e} \gg 1$, we note that the ion contribution to the classical gyrocenter susceptibility \eqref{eq:chi_cl} is dominant if $T_{\rm i} \simeq T_{\rm e}$.

Secondly, the neoclassical bounce-gyrocenter polarization charge density is expressed as
\begin{widetext}
\begin{eqnarray}
\Delta\varrho_{\rm nc} & \equiv & -\sum \int \frac{e^{2}F}{T} \left\langle{\sf T}_{\rm bc} \left( \Phi_{1{\rm bc}} \;-\frac{}{} \langle\Phi_{1{\rm bc}}
\rangle_{\rm b/t}\right)\right\rangle_{\rm b/t} d^{3}\ov{P} = -\sum \frac{e^{2}}{T}\,\ov{\Phi}_{1}e^{i\ov{\Theta}} \int F \left( 1 \;-\frac{}{} \left|\langle e^{i\Delta\Theta}\rangle_{\rm b/t}\right|^{2} \right)\;d^{3}\ov{P},
\label{neoclassical} 
\end{eqnarray}
\end{widetext}
where $d^{3}\ov{P} = 4\pi\,(m\omega_{\|}R_{\|})^{2}B_{0}\,d\mu\,d\kappa/|v_{\|}|$ and we once again assume $F$ to be a Maxwellian distribution. Following Wang and Hahm \cite{Wang_Hahm}, we introduce the flux-surface averaged neoclassical polarization charge density
\begin{eqnarray}
\llbracket\Delta\varrho_{\rm nc}\rrbracket & \equiv & \int_{-\pi}^{\pi}\Delta\varrho_{\rm nc}\,\frac{d\theta}{2\pi} \nonumber \\
 & \equiv &  -\,\sum\;\frac{e^{2}\,N}{T}\,\ov{\Phi}_{1}\,e^{i\ov{\Theta}}\;\chi_{\rm nc},
\label{eq:chi_def}
\end{eqnarray}
and, using the identity 
\[ \left\llbracket \int_{\rm b/t}A\;d\kappa\right\rrbracket \;\equiv\; \int_{\rm b/t}\left\langle \frac{|v_{\|}|\,A}{R_{\|}\omega_{\rm b/t}}\right\rangle _{\rm b/t}\;d\kappa, \]
where we used $\partial\theta/\partial\zeta_{\rm b/t} \equiv |v_{\|}|/(R_{\|}\omega_{\rm b/t})$,  we define the general expression for the flux-averaged neoclassical susceptibility as
\begin{equation}
\chi_{\rm nc} \equiv \sqrt{\frac{2\epsilon}{\pi}} \int\int_{\rm b/t} e^{-\,{\cal E}/T}\,\Gamma_{\rm b/t}(y, \kappa; \epsilon)\;\sqrt{y}\,dy\,d\kappa,
\label{eq:chi_nc_def}
\end{equation}
where we used the definitions $y \equiv \mu\,B_{0}/T$ and ${\cal E}/T = y\,(1 - \epsilon + 2\,\epsilon\kappa)$. In addition, the orbit factor 
$\Gamma_{\rm b/t}$ is defined in the long-wavelength limit $(\alpha \ll 1)$ as
\begin{eqnarray}
\Gamma_{\rm b/t} & \equiv & \left( \begin{array}{c}
1 \\
2
\end{array}\right)\; \left(1 - \left|\left\langle e^{i\Delta\Theta}\right\rangle_{\rm b/t}\right|^{2}\right)\frac{\omega_{\|}}{\omega_{\rm b/t}} 
\label{eq:1_theta_average} \\ 
 & \simeq & \frac{\omega_{\|}\alpha^{2}}{\omega_{\rm b/t}}\;\left\{ \begin{array}{lr}
B_{1}(\kappa) & (0 \leq \kappa < 1) \\
 & \\
2\,\kappa \left[T_{2}(\kappa) - T_{1}^{2}(\kappa)\right] & (\kappa > 1)
\end{array} \right.
\nonumber
\end{eqnarray}
where the additional factor 2 is assigned to the transit-average to account for the contributions from co-passing and counter-passing orbits, and 
\begin{equation}
\alpha \;=\; 2\,\sqrt{\epsilon\,y}\,k_{r}\rho^{\rm T}_{\rm pol}, 
\label{eq:alpha_y}
\end{equation}
with $k_{r}\rho^{\rm T}_{\rm pol}$ defined in terms of the poloidal thermal gyroradius $\rho_{\rm pol}^{T} \equiv 
(T/m\Omega_{\rm pol}^{2})^{1/2} = \rho_{\rm T}\,q/\epsilon \gg \rho_{\rm T}$. Here too, we place ourselves in the long-wavelength approximation $k_{r}\rho^{\rm T}_{\rm pol} \ll 1$.

\subsection{Trapped particles}

\begin{figure}
\epsfysize=2in
\epsfbox{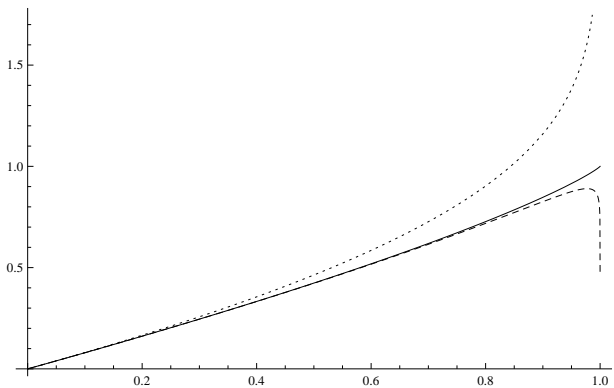}
\caption{Plots of ${\sf K}(\kappa)\,B_{1}(\kappa) = {\sf E}(\kappa) - (1 - \kappa)\;{\sf K}(\kappa)$ (solid), ${\sf K}(\kappa)\,a_{1}^{2}(\kappa)/2$ (dashed), and ${\sf K}(\kappa)\,\kappa/2$ (dotted) in the range $0 < \kappa \leq 1$. The area under each respective curve is $4/9 \simeq 0.444$ (solid), 
$0.437$ (dashed), and $5/9 \simeq 0.556$ (dotted). }
\label{fig:nc_trapped}
\end{figure}

We first consider the trapped-particle contribution (for which $0 \leq \kappa < 1$) to the neoclassical susceptibility \eqref{eq:chi_nc_def}. Using 
Eqs.~\eqref{eq:nu_b_def}, \eqref{eq:B12_def}, and \eqref{eq:1_theta_average}-\eqref{eq:alpha_y}, we find
\begin{eqnarray}
\Gamma_{\rm b} & = & \frac{2\,\alpha^{2}}{\pi}\;{\sf K}(\kappa)\,B_{1}(\kappa) \nonumber \\
 & = & \frac{8\epsilon\,y}{\pi}\,(k_{r}\rho^{\rm T}_{\rm pol})^{2}\; \left[ {\sf E}(\kappa) \;-\frac{}{} (1 - \kappa)\;{\sf K}(\kappa)\right],
\label{eq:Gamma_b}
\end{eqnarray}
so that the trapped-particle contribution to the neoclassical susceptibility \eqref{eq:chi_nc_def} is (to lowest order in $\epsilon$, with 
${\cal E}/T \simeq y$)
\begin{eqnarray}
\chi_{\rm nc}^{\rm tr} & = & 4\,\left(\frac{2\epsilon}{\pi}\right)^{3/2}\,(k_{r}\rho^{\rm T}_{\rm pol})^{2}\times\left(\int_{0}^{\infty}\,y^{3/2}\,
e^{-\,y}\,dy\right) \nonumber \\
 &  &\times\;\left(\int_{0}^{1}\,\left[ {\sf E}(\kappa) \;-\frac{}{} (1 - \kappa)\;{\sf K}(\kappa)\right]\,d\kappa \right) \nonumber \\ 
 & = & 4\,\left(\frac{2\epsilon}{\pi}\right)^{3/2}\,(k_{r}\rho^{\rm T}_{\rm pol})^{2}\;\times\;\frac{3\sqrt{\pi}}{4}\;\times\;
\frac{4}{9} \nonumber \\
 & \simeq & 1.20\;\epsilon^{3/2}\;(k_{r}\rho^{\rm T}_{\rm pol})^{2}.
\label{eq:nc_trapped}
\end{eqnarray}
This expression can be compared to the long-wavelength deeply-trapped neoclassical susceptibility found by Wang and Hahm \cite{Wang_Hahm}, which replaces
$B_{1}(\kappa)$ with $B_{1}(\kappa) \simeq \kappa/2$ in Eq.~\eqref{eq:Gamma_b} and, thus, the factor $4/9$ is replaced with $5/9$ in 
Eq.~\eqref{eq:nc_trapped}. Figure \ref{fig:nc_trapped} shows that the Fourier-Bessel long-wavelength expression \eqref{eq:long_bounce_Bessel} is quite accurate even when $B_{1}(\kappa) \simeq a_{1}^{2}/2$ is taken into account.

\subsection{Passing (circulating) particles}

\begin{figure}
\epsfysize=2in
\epsfbox{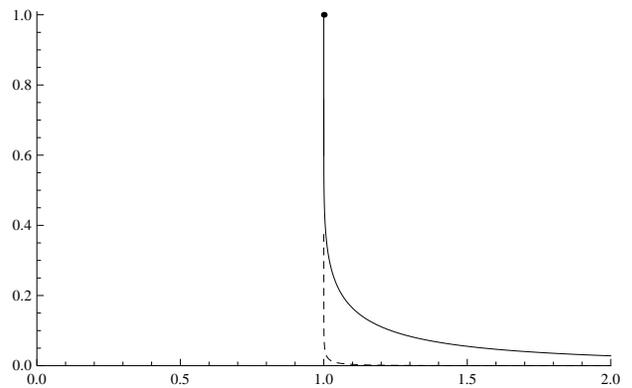}
\caption{Plot of ${\sf K}(\kappa^{-1})\,\sqrt{\kappa}\,[T_{2}(\kappa) - T_{1}^{2}(\kappa)] = \sqrt{\kappa}\,[{\sf E}(\kappa^{-1}) - \pi^{2}/4\,
{\sf K}(\kappa^{-1})]$ in the range $1 \leq \kappa \leq 2$. The dot represents the value of the function at $\kappa = 1$ and the dashed curve represents the difference with the plot of ${\sf K}(\kappa^{-1})\,b_{2}^{2}(\kappa)/(2\sqrt{\kappa})$, which vanishes at $\kappa = 1$. The area under the solid curve is approximately $0.16$.}
\label{fig:nc_passing}
\end{figure}

Next, we consider passing-particle orbits (for which $\kappa > 1$) to the neoclassical susceptibility \eqref{eq:chi_nc_def}. Using 
Eqs.~\eqref{eq:nu_t_def}, \eqref{eq:T_123_def}, and \eqref{eq:1_theta_average}-\eqref{eq:alpha_y}, we find
\begin{eqnarray*} 
\Gamma_{\rm t} & = &  \frac{2\alpha^{2}}{\pi}\;{\sf K}(\kappa^{-1})\,\sqrt{\kappa}\;\left[T_{2}(\kappa) \;-\frac{}{} T_{1}^{2}(\kappa)\right] \\ 
 & = & \frac{8\epsilon\,y}{\pi}\,(k_{r}\rho^{\rm T}_{\rm pol})^{2}\; \sqrt{\kappa}\;\left[ {\sf E}(\kappa^{-1}) \;-\; \frac{\pi^{2}}{4\,{\sf K}(\kappa^{-1})}\right],
\end{eqnarray*}
so that the passing-particle contribution to the neoclassical susceptibility \eqref{eq:chi_nc_def} is  (to lowest order in $\epsilon$, with 
${\cal E}/T \simeq y$)
\begin{eqnarray}
\chi_{\rm nc}^{\rm circ} & = & 4\,\left(\frac{2\epsilon}{\pi}\right)^{3/2}\,(k_{r}\rho^{\rm T}_{\rm pol})^{2}\times\left(\int_{0}^{\infty}\,y^{3/2}\,e^{-\,y}\,dy\right) \nonumber \\
 &  &\times\;\left(\int_{1}^{\infty}\,\sqrt{\kappa}\;\left[ {\sf E}(\kappa^{-1}) \;-\; \frac{\pi^{2}}{4\,{\sf K}(\kappa^{-1})}\right]\;d\kappa \right) \nonumber \\
 & \simeq & 4\,\left(\frac{2\epsilon}{\pi}\right)^{3/2}\,(k_{r}\rho^{\rm T}_{\rm pol})^{2}\times \frac{3}{4}\sqrt{\pi}\;\times\;0.16 \nonumber \\
 & \simeq & 0.43\;\epsilon^{3/2}\;(k_{r}\rho^{\rm T}_{\rm pol})^{2}.
\label{eq:nc_passing}
\end{eqnarray}
We note that the Wang-Hahm result \cite{Wang_Hahm} for the passing-particle contribution replaces the factor $0.43$ with $0.33$, while the long-wavelength limit of the Fourier-Bessel expression 
\begin{equation}
1 - |\langle e^{i\Delta\Theta}\rangle_{\rm t}|^{2} \;=\; 1 - J_{0}^{2}(\alpha\,b_{2}) \;\simeq\; \frac{\alpha^{2}}{2}\,b_{2}^{2}(\kappa) 
\end{equation}
obtained from Eq.~\eqref{eq:transit_approx} is nearly indistinguishable from the curve shown in Fig.~\ref{fig:nc_passing}.

\subsection{Total neoclassical susceptibility}

By adding the trapped/passing-particle contributions \eqref{eq:nc_trapped} and \eqref{eq:nc_passing}, we obtain the total flux-averaged neoclassical susceptibility in the long-wavelength limit:
\begin{eqnarray}
\chi_{\rm nc} & = & \chi_{\rm nc}^{\rm tr} \;+\; \chi_{\rm nc}^{\rm circ} \;\simeq\; 1.63\;\epsilon^{3/2}\;(k_{r}\rho^{\rm T}_{\rm pol})^{2}
\nonumber \\
 & \equiv & 1.63\,\frac{q^{2}}{\sqrt{\epsilon}}\;\chi_{\rm cl},
\label{eq:Chi_nc_final}
\end{eqnarray}
where we used the long-wavelength limit \eqref{eq:chi_cl} of the classical susceptibility $\chi_{\rm cl}$. We note that the Wang-Hahm result 
\cite{Wang_Hahm} replaces the factor $1.63\,(= 1.20 + 0.43)$ with $1.83\,(= 1.50 + 0.33)$ and, therefore, overestimates the trapped-particle contribution (see Fig.~\ref{fig:bounce-coefficient}) and underestimates the passing-particle contribution (see Fig.~\ref{fig:transit_coefficient}) by respectively using the deeply-trapped and strongly-circulating approximations. 

Finally, we note that Eq.~\eqref{eq:Chi_nc_final} is identical to the result obtained by Rosenbluth and Hinton~\cite{ros96} for the flux-surface averaged alpha-particle radial current. The expression for the Rosenbluth-Hinton residual flow~\cite{ros98} is recovered by combining the long-wavelength classical and neoclassical susceptibilities:
\begin{eqnarray}
R_{\mathrm{RH}} & = & \frac{V_{\mathbf{E}\times\mathbf{B}}\left(t\rightarrow\infty\right)}{V_{\mathbf{E}\times\mathbf{B}}\left(t=0\right)}
\;=\; \frac{\chi_{\rm cl}}{\chi_{\rm cl} \;+\; \chi_{\rm nc}} \nonumber \\
 & \simeq & \left(1 \;+\; 1.63\;\frac{q^{2}}{\sqrt{\epsilon}}\right)^{-1}.
\label{eq:RH_result}
\end{eqnarray}
The compact formulas introduced in Sec.~\ref{sec:orbit} could readily be used to generalize this result to include higher-order corrections in powers of 
$\alpha$.

\section{\label{sec:summary}Summary}

By using the elliptic-function representation of the trapped/passing-particle guiding-center orbits in simple axisymmetric tokamak geometry (presented in Sec.~\ref{sec:gc_orbits}), we have derived explicit compact expressions for the orbit-averaged eikonal factor $\langle e^{i\Delta\Theta}\rangle_{\cal O}$ that appears in bounce-gyrokinetic theory. 

For the case of the bounce average associated with trapped-particle guiding-center orbits, we obtained
\begin{eqnarray}
\langle e^{i\Delta\Theta}\rangle_{\rm b} & = & \sum_{m=0}^{\infty}\;\frac{(-1)^{m}\,\alpha^{2m}}{(2m)!}\;B_{m}(\kappa) \nonumber \\
 & \simeq & J_{0}\left(\alpha\frac{}{}a_{1}(\kappa)\right),
\end{eqnarray}
where the coefficients $B_{m}(\kappa)$ are defined in Eqs.~\eqref{eq:B_def}-\eqref{eq:B12_def} while the coefficient $a_{1}(\kappa)$ is defined in 
Eq.~\eqref{eq:a_def}. For the case of the transit average associated with passing-particle guiding-center orbits, we obtained
\begin{eqnarray}
\langle e^{i\Delta\Theta}\rangle_{\rm t} & = & \sum_{n=0}^{\infty}\;\frac{(i\sigma\alpha\sqrt{\kappa})^{n}}{n!}\;T_{n}(\kappa) \nonumber \\
 & \simeq & e^{i\sigma\alpha\nu_{\rm t}(\kappa)/2}\;J_{0}\left(\alpha\frac{}{}b_{2}(\kappa)\right),
\end{eqnarray}
where the coefficients $T_{n}(\kappa)$ are defined in Eqs.~\eqref{eq:T_def}-\eqref{eq:T_123_def} while the coefficients $\nu_{\rm t}(\kappa)$ and 
$b_{2}(\kappa)$ are defined in Eqs.~\eqref{eq:nu_t_def} and \eqref{eq:b_def}. These compact formulas were used to recover the Rosenbluth-Hinton result \eqref{eq:RH_result} more accurately than previously calculated by Wang and Hahm \cite{Wang_Hahm}. Therefore, our results are applicable to problems in which an expression for the neoclassical susceptibility \eqref{eq:chi_nc_def} is valid over a wide range of wavelengths is needed \cite{Hahm_13}, where 
Eq.~\eqref{eq:1_theta_average} is expressed as
\begin{equation}
\Gamma_{\rm b/t} \;=\; \left( \begin{array}{c}
1 \\
2
\end{array}\right)\frac{\omega_{\|}}{\omega_{\rm b/t}} \left\{ \begin{array}{lr}
1 - J_{0}^{2}(\alpha a_{1}) & (0 \leq \kappa < 1) \\
 & \\
1 - J_{0}^{2}(\alpha b_{2}) & (\kappa > 1)
\end{array} \right.
\end{equation}
and the coefficients $a_{1}(\kappa)$ and $b_{2}(\kappa)$ are defined in Eqs.~\eqref{eq:a_def} and \eqref{eq:b_def}, respectively.

Lastly, the compact formulas presented here can be extended to the case of self-consistent bounce-gyrokinetic Vlasov-Maxwell equations, where bounce-gyrocenter magnetization associated with magnetic-field fluctuations in simple axisymmetric tokamak geometry can also be calculated explicitly.

\acknowledgments

Work by FXD and TSH was supported by the Brain Korea 21 Plus Project (No.~21A20130012821) of Korea, by the World Class Institute (WCI) Program of the National Research Foundation of Korea (NRF) funded by the Ministry of Science, ICT \& Future Planning (MSIP) (No. WCI-2009-0001) and by the National R\&D Program through the National Research Foundation of Korea (NRF) funded by the Ministry of Science, ICT \& Future Planning (MSIP) (No. 2012M1A7A1A02034). Work by AJB was supported by a U.S.~DoE grant under contract No.~DE-SC0006721.


\begin{thebibliography}{5}

\bibitem{Brizard_Hahm} A.~J.~Brizard and T.~S.~Hahm, Rev.~Mod.~Phys.~{\bf 79}, 421 (2007).

\bibitem{Lin} Z.~Lin, T.~S.~Hahm, W.~W.~Lee, W.~M.~Tang, and R.~B.~White, Science {\bf 281}, 1835 (1998). 

\bibitem{Garbet} X.~Garbet, Y.~Idomura, L.~Villard, and T.~H.~Watanabe, Nucl.~Fusion {\bf 50}, 043002 (2010).

\bibitem{Brizard_2013} A.~J.~Brizard, Phys.~Plasmas {\bf 20}, 092309 (2013).

\bibitem{Brizard_2008} A. J. Brizard, Commun.~Nonlinear Sci.~Numer.~Simul.~{\bf 13}, 24 (2008).

\bibitem{Fong_Hahm} B.~H.~Fong and T.~S.~Hahm, Phys.~Plasmas {\bf 6}, 188 (1999).

\bibitem{Brizard_2000} A.~J.~Brizard, Phys.~Plasmas {\bf 7}, 3238 (2000).

\bibitem{Wang_Hahm} L.~Wang and T.~S.~Hahm, Phys.~Plasmas {\bf 16}, 062309 (2009).

\bibitem{Brizard_2011} A.~J.~Brizard, Phys.~Plasmas {\bf 18}, 022508 (2011).

\bibitem{Brizard_2014} A.~J.~Brizard and F.-X.~Duthoit, Phys.~Plasmas {\bf 21}, 052509 (2014).

\bibitem{NIST_Jacobi} W.~P.~Reinhardt and P.~L.~Walker, {\it Jacobian Elliptic Functions}, in NIST Handbook of Mathematical Functions (Cambridge University Press, Cambridge, 2010), Chap.~22.

\bibitem{Lawden} D.~F.~Lawden, {\it Elliptic Functions and Applications}, (Springer-Verlag, New York, 1989).

\bibitem{ros98} M.~N.~Rosenbluth and F.~L.~Hinton, Phys.~Rev.~Lett. \textbf{80}, 724 (1998).

\bibitem{Zonca} F.~Zonca, P.~Buratti, A.~Cardinali, L.~Chen, J.-Q.~Dong, Y.-X.~Long, A.~V.~Milovanov, F.~Romanelli, P.~Smeulders, L.~Wang, Z.-T.~Wang, 
C.~Castaldo, R.~Cesario, E.~Giovannozzi, M.~Marinucci, and V.~Pericoli Ridolfini, Nucl.~Fusion {\bf 47}, 1588 (2007).

\bibitem{footnote} We note that our definition of $\kappa$, which is consistent with the $\kappa$-dependence of elliptic functions and integrals in {\sf Mathematica}, corresponds to the square of $\kappa_{WH}$ used by Wang and Hahm \cite{Wang_Hahm}.

\bibitem{RGL_82} R.~G.~Littlejohn, Phys.~Scr.~{\bf T1/2}, 119 (1982).

\bibitem{Cary_Brizard} J.~R.~Cary and A.~J.~Brizard, Rev.~Mod.~Phys.~{\bf 81}, 693 (2009)

\bibitem{Gang_Diamond} F.~Y.~Gang and P.~H.~Diamond, Phys.~Fluids B {\bf 2}, 2976 (1990).

\bibitem{dat92}G.~Dattoli, C.~Chiccoli, S.~Lorenzutta, G.~Maino, M.~Richetta, and A.~Torre, J. Math. Phys. \textbf{33}, 25 (1992)

\bibitem{dat98}G.~Dattoli, C.~Chiccoli, A.~Torre and S.~Lorenzutta, Le Matematiche \textbf{53}, 387 (1998)

\bibitem{Brizard_reduced_1} A.~J.~Brizard, Comm.~Nonlin.~Sci.~Num.~Sim.~{\bf 13}, 24 (2008).

\bibitem{Brizard_reduced_2} A.~J.~Brizard, J.~Phys.: Conf.~Ser.~{\bf 169}, 012003 (2009).

\bibitem{Brizard_reduced_3} A.~J.~Brizard, Phys.~Plasmas {\bf 20}, 092309 (2013).

\bibitem{Hahm_13} T.~S.~Hahm, L.~Wang, W.~X.~Wang, E.~S.~Yoon, and F.-X.~Duthoit, Nucl.~Fusion {\bf 53}, 072002 (2013).


\end{thebibliography}
\end{document}